# Analysis of Wireless Power Transmission

A. Mahmood[1], H. Fakhar[1], S. H. Ahmed[2], N. Javaid[1,*]

COMSATS Institute of Information Technology, Islamabad Pakistan.
School of Computer Science & Engineering, Kyungpook National University, Korea.

*Abstract-* **The advent of various wireless technologies have revolutionized the communication infrastructure and consequently changed the entire world into a global village. Use of wireless technology has also been made for transmission of electric power wirelessly. It increases the portability of power systems and integrates the communication technologies and electric power to the same platform. This paper presents a comprehensive review and detailed analysis of various techniques used for wireless power transmission. Feasibility, implementations, operations, results and comparison among different methods have also been covered in order to identify the favorable and economical method for low power and small distance applications.**

I. INTRODUCTION

Goal of wireless power transmission is to transfer electricity from source to destination in order to reduce high transmission losses. Nicola Tesla is the pioneer of the wireless power transmission concept and he implemented his concept in 1902 with the technologies available at that time. Since then, many of the scientists involved in further extension of this idea and achieved positive results for the transmission at a close range [1]. In 2007, a team at the Massachusetts Institute of Technology (MIT), was successful in transferring the power wirelessly at a mid range using magnetic coupling resonance by lighting a bulb 0f 60w at a range of 2m [2]. Wireless power transmission increases the portability and convenience. It also fulfils the demand of recent devices and technologies which already use wireless technique for different operations and communications like Wi-Fi being used in laptops and cellular phones for the access to the internet. Transfer of electric power without using wires is possible by using three major methods. The first method is to transfer electric power by the phenomena of mutual induction between two coils operating at same resonant frequency; second method is by microwave transmitter and receiver while the third method is the transfer of electric power using laser technology [3][4].

Rest of the paper is organized as follows. Section II is dedicated for microwave power transmission and section III is about laser transmission technique. While section IV describes the magnetic resonance method followed a comparison of these three methods. Section V elaborates the advantages and applications associated with wireless power transmission. Conclusions are drawn in section VI.

II. WIRELESS POWER TRANSMISSION USING MICROWAVES

Microwave wireless power transmission is a wide range process in which long distance electric power transmission becomes possible. This process uses the microwave voltage source which emits the microwaves. The microwave source acts as a transmitting antenna and a microwave receiver is attached with the load which acts as receiving antenna. The received microwaves are then converted back in to electrical energy through which the load is driven. Different parts of the wireless power transmission through microwaves are briefed as following.

The microwave source antenna acts as transmitting antenna at the base station. It uses the high frequency microwaves ranging from 1GHz to 1000 GHz. There are many types of microwaves source antennas having different efficiencies. Usually the slotted wave guide, micro strip patch and parabolic dish antennas are used for this purpose. For high power applications the slotted waveguide antennas are used because of their high efficiency. The microwave receiving antenna is mounted at the load end and due to high frequency of microwaves it could be used for large distance applications of wireless power transmission. The unit which receives microwaves and then converts back to the dc power is called rectenna.

III. WIRELESS POWER TRANSMISSION USING LASER

The second technique used for wireless power transmission is based on laser beam which acts as a source. The laser beam of high intensity is thrown from some specific distance to the load end. Depending on the range and intensity of the beam this method is used for small distance applications. This process is similar to the solar cells photovoltaic generation which uses the solar energy of the sun light and converts it to electricity. At the load end highly efficient photo voltaic cells are used which receive the laser beam, energize laser light and finally convert light energy into electrical energy.

Experiments have shown that the wireless power transmission through laser beam is 50 percent efficient with respect to other methods but by using advance technology of laser photovoltaic cell receivers; the efficiency could be increased [5]. The laser source transmits the laser beam through an efficient lens. The lens is used to converge the beam of the laser to the specific place where the receiver is present. The load is attached with the photovoltaic cells which after being energized through laser beam convert light energy of laser beam into electrical energy.

IV. WIRELESS POWER TRANSMISSION BY MAGNETIC RESONANCE

The mutual induction phenomena states that if there is a continuous varying current passes through one coil produces the magnetic field in the space around first coil called primary coil. As this varying magnetic field interacts with the secondary coil it produces an induced current in the secondary coil. This is also called magnetic resonance between two coils operating at a same resonance frequency. Principle of mutual induction is elaborated in Fig. 1.

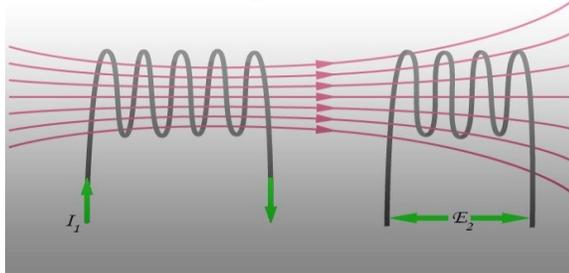

Fig. 1. Mutual induction process

The frequency at which the amplitude of the waves produced in the system is maximum called resonance frequency. The resonance frequency is attained by varying different parameters affecting the gain of voltage produced within the coils. The phenomena of wireless power transmission using mutual induction consist of two coils known as primary and secondary coils. These coils act as transmitting and receiving antennas. The process is described in block diagram in Fig. 2.

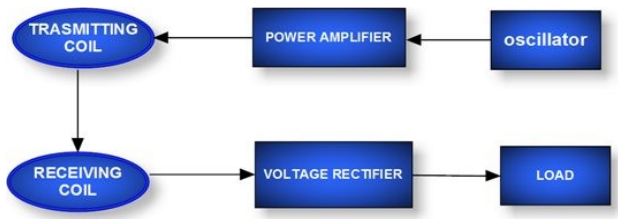

Fig. 2. Magnetic resonance method

A set of copper wires is used which acts as transmitting and receiving coil. The number of turns of copper wires, resistance and diameter affects the mutual induction between them. An oscillator is used for generation of a particular voltage signal of such a frequency where the mutual induction will be maximum. Such frequency is called resonance frequency. Same process could be done by using signal generator. There are many types of oscillators. Voltage control oscillator is one of them. Above mentioned methods are compared in Table 1.

TABLE I
COMPARISON OF DIFFERENT METHODS

| Magnetic Resonance Method | Microwaves Method | Lasers Method |
| --- | --- | --- |
| It is economical as the equipment used is cheap and easily available | Relatively expensive as compared to other methods | Implies same economic conditions of mutual induction |
| Useful for implementation of the small distance applications | This method implies for long distance applications | Used for small distance but could be used for longer distances (high intensity) |
| It is safe from biological point of view. | Injurious for health because of high frequency rays | The laser method is also injurious to human health |

## V. ADVANTAGES AND APPLICATIONS

Wireless power transmission is the only process through which we could eliminate the existing system of high power transmission lines, towers and substations, which are seen as not very efficient way of energy transmission and require a huge amount to be maintained time to time. This will lead to a globally efficient and cheap transmission system. The cost of the transmission and receiving power would lower for the daily users and the large scale reduction of power tariff would be easily visible. The loss of transmission will be decreased and the power could easily be transferred to any place irrespective of the geographical situations. The power failures would be minimized which occur due to short circuit or fault in cable lines, making it a more efficient and environment friendly system.

Portability, convenience and the demand of recent developing technologies can be fulfilled by wireless power transmission. It have vast applications which decrease the cost of power systems by elimination of wires and towers thus saving the cost of equipments and labor, decreasing the complexity and increasing the efficiency of a power system.

## VI. CONCLUSIONS

The wireless power transmission is indeed a great and a noble idea. In this paper we have reviewed and compared different wireless power transmission methods. Different phenomenon governing this idea along with equipment has been explored. Advantages and applications have been elaborated.

*Corresponding Author: email: nadeemjavaid@comsats.edu.pk, nadeemjavaid@yahoo.com, web: www.njavaid.com